\title{Quant Trefoil}
\author{ivanviktd }
\date{December 2021}

\documentclass{article}
\pdfoutput=1
\usepackage{lscape}
\usepackage{enumerate}
\usepackage{amsfonts}
\usepackage{amsmath}
\usepackage{comment}
\usepackage{textcomp}
\usepackage{amsthm, tikz}
\usepackage{amssymb}
\usepackage{color}
\usepackage{graphicx}
\usepackage{pdflscape}
\usepackage{afterpage}
\usepackage{array}

\hoffset= - 1.2in         

\begin{document}

\title{\vspace{0.1cm}{\Large {\bf Calculating HOMFLY-PT polynomials on a photonic processor}\vspace{.2cm}}
\author{\bf Ivan Dyakonov$^{a,b}$\thanks{e-mail: iv.dyakonov@physics.msu.ru}, Ilya Kondratyev$^{a}$, Sergey Mironov$^{c,d,e}$\thanks{e-mail: sa.mironov\_1@physics.msu.ru},
Andrey Morozov$^{d,f,g}$\thanks{e-mail: morozov.andrey.a@iitp.ru}}
}
\date{ }

\maketitle

\vspace{-5.5cm}

\begin{center}
\hfill ITEP/TH-13/24\\
\hfill IITP/TH-11/24\\
\hfill MIPT/TH-9/24\\
\end{center}

\vspace{3.6cm}

\begin{center}
$^a$ {\small {\it Quantum Technology Centre, Faculty of Physics, MSU, Moscow, 119991, Russia}}\\
$^b$ {\small {\it Russian Quantum Center, Moscow, 121205, Russia}}\\
$^c$ {\small {\it INR RAS, Moscow 117312, Russia}}\\
$^d$ {\small {\it NRC ``Kurchatov Institute'', Moscow 123182, Russia}}\\
$^e$ {\small {\it ITMP, MSU, Moscow, 119991, Russia}}\\
$^f$ {\small {\it IITP RAS, Moscow 127994, Russia}}\\
$^g$ {\small {\it MIPT, Dolgoprudny, 141701, Russia}}\\
\end{center}

\vspace{1cm}

\begin{abstract}
In this paper we discuss an approach to calculate knot polynomials on a photonic processor. Calculations of knot polynomials is a computationally difficult problem and therefore it is interesting to use new advanced calculation methods to find them. Here we present a proof of concept by calculating the simplest knot polynomial of the trefoil knot in fundamental representation. This approach, however, can easily be generalized to more complex knots and representations. Same operators can also be realized on a quantum computer with the same effect.

\end{abstract}


\vspace{.5cm}



\section{Introduction}

Coherent photonic processors have recently attracted attention due to capability to significantly improve speed and energy efficiency of executing a particular kind of operations - unitary matrix multiplication \cite{Zhang2022}. The principle of operation is simple: a multimode input light field $a^{in}_{i}$ undergoes a transformation $U$ imposed by an interferometric circuit. The output field is then expressed as $a_{j}^{out}=U_{ij}a^{in}_{i}$. The advance of integrated photonic technologies enabled fabrication of precisely reconfigurable universal interferometers \cite{Harris18, Taballione2023} that provide full control over the unitary matrix $U$ via the phase modulation inside interferometer arms. Reconfigurable photonic processors find wide application in boosting routine tasks of machine learning algorithms \cite{Englund2017} and in experimental simulation of fundamental physics models \cite{vanDerMeer2023}. The perspective of higher signal processing bandwidth and reduced energy consumption provides steady development of integrated photonics \cite{Chrostowski2024}.

We want to apply coherent photonic processor to calculations in mathematical physics. The calculations we are interested in in the context of this paper are the calculations of the knot polynomials or Wilson-loop averages of the Chern-Simons theory.

Chern-Simons theory with the action
\begin{equation}
S=\frac{k}{4\pi}\int \mathcal{A}\wedge d\mathcal{A}+\frac{2}{3}\mathcal{A}\wedge\mathcal{A}\wedge\mathcal{A}=
\frac{k}{4\pi}\int d^3x \varepsilon^{ijk}\left(\delta_{ab}\mathcal{A}^a_i\partial_j\mathcal{A}^b_k+\frac{2}{3}if_{abc}\mathcal{A}^a_i\mathcal{A}^b_j\mathcal{A}^c_k\right).
\end{equation}
 is usually considered in the literature as potential model for topological quantum computer as it describes non-Abelian anyons. Though we do not have such physical system yet, topological quantum computer would have radically better fidelity in comparison to many other quantum computer models as the states are topologically stable \cite{tqc}.

In this paper though, we consider the opposite problem. As Chern-Simons theory itself when employed in knot theory presents technical interest, we suggest to use an existing quantum computer to calculate interesting quantities in such a theory and, consequently, in topological theory of knots.
In Chern-Simons theory the most interesting and complex observables are Wilson-loop averages:
\begin{equation}
\left<W\right>_{Q}^{\mathcal{K}}=\text{Tr}_{Q}\text{ Pexp}(\int\limits_{\mathcal{K}}\mathcal{A}dx),
\end{equation}
here $\mathcal{K}$ is a contour in 3d space along which we calculate Wilson-loop, which is in fact a knot, $T$ is a representation of the gauge group in which we calculate the Wilson-loop averages. These observables, as it is known \cite{Witt}, are equal to the knot polynomials from the mathematical knot theory, with the knot being equivalent to the loop $\mathcal{K}$. These knot polynomials are interesting from the point of view of many theories because of numerous dualities with other theories, for example conformal field theory \cite{Witt}-\cite{WZNWend}, topological string theory \cite{topstr}, integrable theories \cite{int} and matrix models \cite{mm1}-\cite{mm4} and others. In the context of topological quantum computers, knot polynomials are potential observables. Due to the fact that the knot polynomials on one hand are constructed from a product of unitary matrices and, on the other hand, their calculations prove to be a difficult task for complex knots, calculations of knot polynomials is  a natural problem for the advanced computing machines, such as quantum computer and photonic processor.

We study Chern-Simons theory with the gauge group $sl_N$ and calculate the resulting knot polynomials which are called HOMFLY-PT polynomials \cite{HOMFLY,PT} on a photonic processor. HOMFLY-PT are polynomials of two variables $q$ and $A$ which are connected to the parameters of the theory
\begin{equation}
q=exp(\frac{4\pi i}{k+N}),\ \ \ A=q^N.
\label{eq:Aqdef}
\end{equation}

\section{Two-bridge knot calculus \label{secTh}}

\begin{figure}[h!]
\begin{picture}(200,100)(-255,-55)
\qbezier(0,0)(-12,12)(-24,12)
\qbezier(-24,12)(-36,12)(-46,2)
\qbezier(-50,-2)(-60,-12)(-72,-12)
\qbezier(-72,-12)(-84,-12)(-84,-24)
\qbezier(0,0)(12,-12)(24,-12)
\qbezier(24,-12)(36,-12)(46,-2)
\qbezier(50,2)(60,12)(72,12)
\qbezier(72,12)(84,12)(84,24)
\qbezier(2,2)(12,12)(24,12)
\qbezier(24,12)(36,12)(48,0)
\qbezier(48,0)(60,-12)(72,-12)
\qbezier(72,-12)(84,-12)(84,-24)
\qbezier(-2,-2)(-12,-12)(-24,-12)
\qbezier(-24,-12)(-36,-12)(-48,0)
\qbezier(-48,0)(-60,12)(-72,12)
\qbezier(-72,12)(-84,12)(-84,24)
\qbezier(84,24)(84,36)(72,36)
\qbezier(84,-24)(84,-36)(72,-36)
\qbezier(-84,24)(-84,36)(-72,36)
\qbezier(-84,-24)(-84,-36)(-72,-36)
\put(-72,36){\line(1,0){144}}
\put(-72,-36){\line(1,0){144}}
\end{picture}
\caption{Trefoil knot as a two-bridge knot.\label{F2br}}
\end{figure}
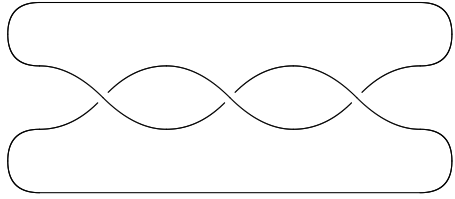

We adopt the Reshetikhin-Turaev approach to knot calculus \cite{RT}-\cite{RTend} and use a simple construction of the two-bridge knot. Such knots are constructed as a four strand braid (see Fig.\ref{F2br}), closed on each side.

The polynomials are constructed as matrix elements of the product of operators that correspond to each crossing in the knot \cite{2br}-\cite{2brend}. Moreover, crossing-operators between different strands are connected via Racah matrices. Important property of all these operators is that they are represented by the unitary matrices for $q$ from (\ref{eq:Aqdef}) and large enough $k$. Such matrices can also be used as universal quantum gates in topological quantum computer \cite{Rgates,Rgates2,Rgates3}.

Knot polynomials and operators depend on the representation of the group $sl_N$. Chosen representation define the size of matrices and thus change the size of quantum computer required to calculate corresponding knot polynomial. For the simplest fundamental representation that we consider in this paper the size of matrices is $2$ by $2$ and thus one-qubit quantum computer is needed. For the fundamental representation required operators are equal to


\begin{equation}
\begin{array}{lcrlcclcrl}
&&\begin{array}{cc}[2]\ \ \ \ \ \ \ \ &[1,1]\ \ \ \ \ \ \ \end{array}
&&&&&\begin{array}{cc}[2]\ \ \ &[1,1]\ \ \ \end{array}\\
\\ \\
S_{[1]}&=&
\frac{1}{\sqrt{(q+q^{-1})(A-A^{-1})}}\left(\begin{array}{cc}\sqrt{\frac{A}{q}-\frac{q}{A}} & \sqrt{Aq-\frac{1}{Aq}} \\ \sqrt{Aq-\frac{1}{Aq}} & -\sqrt{\frac{A}{q}-\frac{q}{A}} \end{array}\right)
& \begin{array}{l}\,\emptyset\\ \, \text{adj}\end{array}
&&
T_{[1]}&=&
\left(\begin{array}{cc}\frac{q}{A} & 0 \\ 0 & -\frac{1}{qA} \end{array}\right)
& \begin{array}{l}\,[2]\\ \, [1,1]\end{array}
\\ \\ \\ \\
&&\begin{array}{lc}\emptyset\ \ \ \ \ \ \ \ \ \ \ \ \ \ \ \ \ \ \ \ &\text{adj}\ \ \ \ \ \ \ \ \ \ \ \end{array}
&&&&&\begin{array}{lc}\emptyset\ \ &\text{adj}\ \ \end{array}
\\ \\
\bar{S}_{[1]}&=&
\left(\begin{array}{cc}\frac{q-q^{-1}}{A-A^{-1}} & \frac{\sqrt{(Aq-\frac{1}{Aq})(\frac{A}{q}-\frac{q}{A})}}{A-A^{-1}} \\ \frac{\sqrt{(Aq-\frac{1}{Aq})(\frac{A}{q}-\frac{q}{A})}}{A-A^{-1}} & -\frac{q-q^{-1}}{A-A^{-1}} \end{array}\right)
& \begin{array}{l}\,\emptyset\\ \, \text{adj}\end{array}
&&
\bar{T}_{[1]}&=&
\left(\begin{array}{cc} 1 & 0 \\ 0 & -A \end{array}\right)
& \begin{array}{l}\,\emptyset\\ \, \text{adj}\end{array}
\\ \\
\end{array}
\end{equation}

The knot polynomial is then given by the matrix element of the product of aforementioned operators. For the simplest trefoil knot the answer is given by the following formula (see also Fig.\ref{F2br}):
\begin{equation}
\mathcal{H}_{[1]}^{3_1}=\frac{A-A^{-1}}{q-q^{-1}}\left(S T^3 S^{\dagger}\right)_{\emptyset\emptyset}=A^2-A^4(q^2+q^{-2})
\end{equation}
Actually the simplest way to calculate this polynomial is to use non-diagonal operators:
\begin{equation}
T_{nd}=STS^{\dagger}=\left(\begin{array}{cc}
-A^2\cfrac{q-q^{-1}}{A-A^{-1}}& A\cfrac{\sqrt{(Aq-A^{-1}q^{-1})(A/q-q/A)}}{A-A^{-1}}
\\ \\
A\cfrac{\sqrt{(Aq-A^{-1}q^{-1})(A/q-q/A)}}{A-A^{-1}} & \cfrac{q-q^{-1}}{A-A^{-1}}
\end{array}\right)
\label{eq:calc}
\end{equation}
Then the trefoil polynomial is a matrix element of $T_{nd}^3$. In general powers of $T_{nd}$ will give different two-strand knots and links.

\section{Experiment and results}

Coherent photonic processors have recently drawn attention due to ability to perform computation of specific mathematical functions effectively in terms of time and energy. The most vivid example is matrix multiplication implemented simply by transmitting light through an optical interferometer and detecting output power. 
In this work we devised a method to map the problem of computing the knot polynomial to a coherent photonic platform.

\begin{figure}[h!]
\centering
\includegraphics[width=10cm]{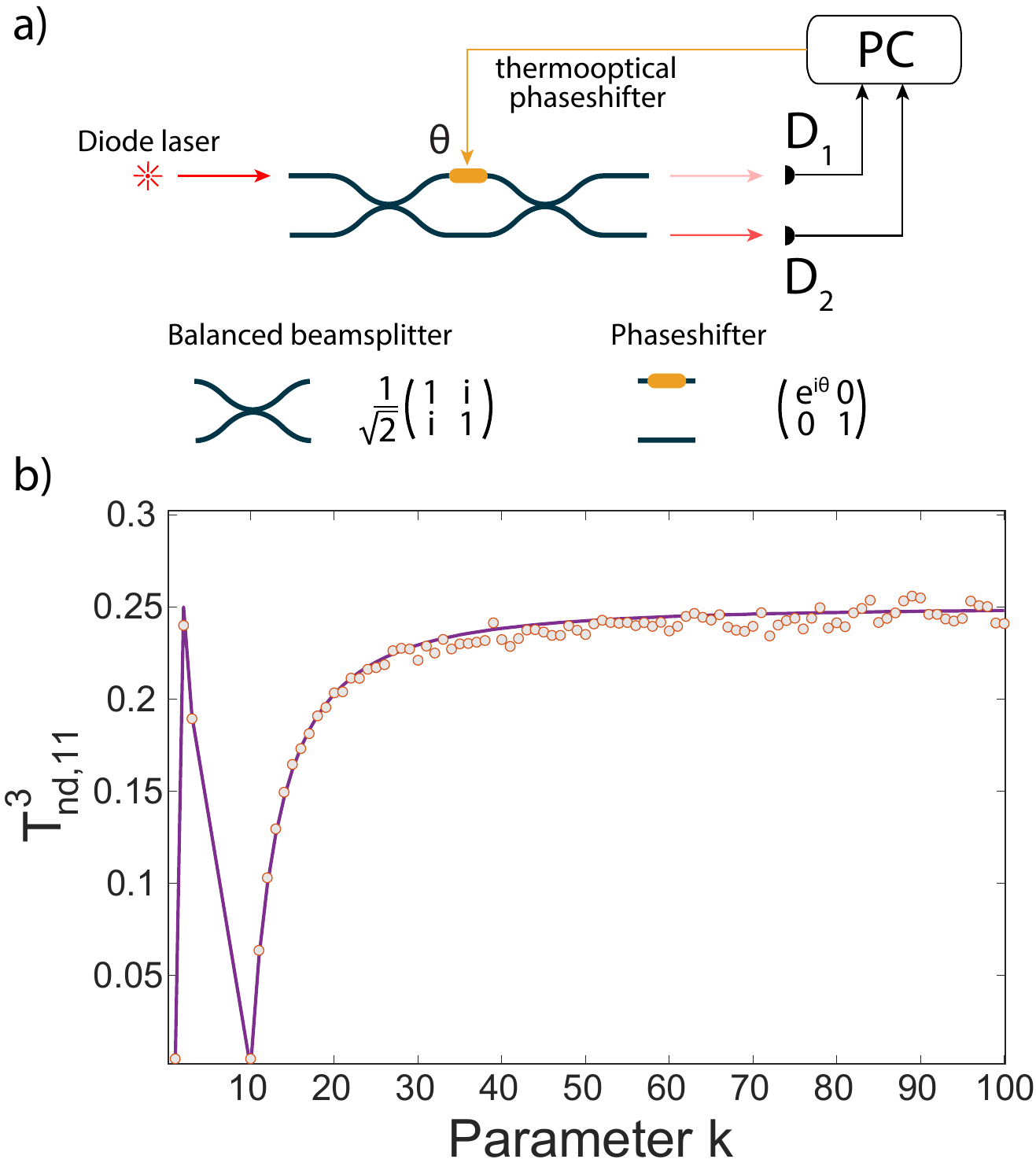}
\caption{a) Schematic of the photonic processor. The emission of a diode laser is coupled into one of the input ports of the integrated Mach-Zehnder interferometer. The relative phase $\theta$ between the arms is controlled by a thermooptical phaseshifter, driven by a programmable current source connect to a personal computer. The output power distribution is registered by a pair of detectors $D_{1}$ and $D_{2}$. The measured values are normalized to a sum of output power in both ports. b) The result of the $T^{3}_{nd,11}$ measurement. The value of the matrix element is estimated as the normalized output power, measured by the detector $D_{1}$. The dots represent the experimental results, the solid violet curve represents theory.\label{fig:photonic_processor_scheme}}
\end{figure}

Our photonic processor implements multiplication of $2\times2$ unitary matrix and a two-dimensional complex field amplitude vector. The schematic of the processor can be found in Fig. \ref{fig:photonic_processor_scheme}. We use the femtosecond laser writing method to fabricate a reconfigurable Mach-Zehnder interferometer. The photonic waveguides are inscribed by a tightly focused beam of the ytterbium-fiber femtosecond laser (Avesta Antaus) inside the volume of a fused silica sample mounted on the 3D air-bearing stage (Aerotech FiberGlide 3D). The phase modulation of one the interferometer arms is implemented using a thermo-optic phaseshifter. The phaseshifer is engraved on top of the photonic circuit after the NiCr metal film is sputtered onto the top surface of the chip. The electrical connection is established via spring-loaded electric contacts soldered onto the interface PCB board. We drive the circuit with constant current set by a programmable current source.

The input light field is generated by a diode laser (Toptica CTL 950) and injected into port X of the circuit through a single-mode fiber array. The output of the chip is coupled to another single-mode fiber array which is connected to a pair of PIN-diode-based photodetectors which measure the output power.

Using photonic processor, we have implemented (\ref{eq:calc}). Then we measured the absolute value of the corresponding matrix elements for different values of $q$ (different $k$). As a result we get graph on Fig. \ref{fig:photonic_processor_scheme}. On this graph we present experimental measurements, depending on the values of $k$. As we can see it coincides with theoretical curve, which shows that we can indeed measure knot polynomials, using this approach.


\section{Conclusion}

This paper should be considered as a \textit{proof-of-concept} for calculating knot polynomials on a quantum computer or photonic processor. We used the simplest case of fundamental representation for the 2-bridge knots 
{which can be calculated using a one qubit quantum computer or similar photonic processor}. We show that corresponding matrix element calculated on a photonic processor indeed reproduces the analytical answer. This approach should be also expanded to higher representations and more involved knots which would require 
higher dimensional photonic processors or multiqubit quantum computers.
%

\begin{thebibliography}{99}

\bibitem{Zhang2022} Z.~Hailong and D.~Jianji and C.~Junwei and D.~Wenchan and H.~Chaoran and S.~Yichen and Z.~Qiming and G.~Min and Q.~Chao and C.~Hongsheng and R.~Zhichao and Z.~Xinliang,
     Light Sci Appl 11, 30 (2022).

\bibitem{Harris18}  N.~C.~Harris, J.~Carolan, D.~Bunandar, M.~Prabhu, M.~Hochberg, T.~Baehr-Jones, M.~L.~Fanto, A.~Matthew Smith, C.~C.~Tison, P.~M.~Alsing and D.~Englund,
    Optica 12 (2018) vol.5 1623--1631

\bibitem{Taballione2023} C.~Taballione, M.~Correa Anguita, M.~de Goede, P.~Venderbosch, B.~Kassenberg, H.~Snijders, N.~Kannan, W.~L.~Vleeshouwers, D.~Smith and J.~P.~Epping, \textit{et al.}
Quantum \textbf{7} (2023), 1071, arXiv:2203.01801.

\bibitem{Englund2017} Y.~Shen, N.~Harris, S.~Skirlo, \textit{et al.} Nature Photon 11, 441–446 (2017)

\bibitem{vanDerMeer2023} R.~van der Meer, Z.~Huang, M.~C.~Anguita, \textit{et al.} npj Quantum Inf 9, 32 (2023)

\bibitem{Chrostowski2024} S.~Shekhar, W.~Bogaerts, L.~Chrostowski, \textit{et al.} Nat Commun 15, 751 (2024)

\bibitem{tqc} A. Kitaev,
Annals of Physics. 303 (2003) 2–30, arXiv:quant-ph/9707021v1

\bibitem{Witt} E. Witten,
Commun. Math. Phys.121 (1989) 351-399

\bibitem{WZNW1} Kaul R.K.,
Commun. Math. Phys. 162 (1994) 289–320, arXiv:hep-th/9305032

\bibitem{WZNW2} Kaul R.K., Govindarajan T.R.,
Nucl. Phys. B380 (1992) 293–336, arXiv:hep-th/9111063

\bibitem{WZNW3} Ramadevi P., Govindarajan T.R., Kaul R.K.,
Nucl. Phys. B402 (1993) 548–566, arXiv:hep-th/9212110

\bibitem{WZNW4} Ramadevi P., Govindarajan T.R., Kaul R.K.,
Nucl. Phys. B422 (1994) 291–306, arXiv:hep-th/9312215

\bibitem{WZNW5} Kaul R.K.,
Frontiers of field theory, quantum gravity and strings. Proceedings (1999) 45–63, arXiv:hep-th/9804122

\bibitem{WZNW6} Guadagnini E., Martellini M., Mintchev M.,
Phys. Lett. B227 (1989) 111–117

\bibitem{WZNW7} Guadagnini E., Martellini M., Mintchev M.,
Nucl. Phys. B330 (1990) 575–607

\bibitem{WZNW8} Alvarez M., Labastida J.M.F.,
Nucl. Phys. B395 (1993) 198–238, arXiv:hep-th/9110069

\bibitem{WZNW9} Axelrod S., Singer I.M.,
Proc. of XXth DGM conference.— New York : World Scientific  (1991) 3–45, arXiv:hep-th/9110056

\bibitem{WZNW10} Labastida J.M.F., Ramallo A.V.,
Phys. Lett. B228 (1989) no.2

\bibitem{WZNWend} Frohlich J., King C.,
Commun. Math. Phys. 126 (1989) 167–199

\bibitem{topstr} Ekholm T.,
arXiv:1312.0800

\bibitem{int} Mironov A., Morozov A., Morozov And.,
Strings, gauge fields, and the geometry behind: the legacy of Maximilian Kreuzer.— Singapore : World Scientific, (2013)  101–118,  arXiv:hep-th/1112.5754

\bibitem{mm1} Tierz M.,
Mod. Phys. Lett. A19 (2004) 1365–78

\bibitem{mm2}Brini A., Eynard B. and Marino M.,
Ann. Henri Poincaré 13 (2012) 1873–910

\bibitem{mm3} Alexandrov A., Mironov A., Morozov A. and Morozov An., 
J. Exp. Theor. Phys. Lett. 100 (2014) 297–304, arXiv:1407.3754

\bibitem{mm4} Alexandrov A. and Melnikov D., 
arXiv:1411.5698

\bibitem{HOMFLY}  Hoste J., Ocneanu A., Millett K., Freyd P., Lickorish W.B.R., Yetter D.,
    Bull. Amer. Math. Soc. 12 (1985) 239-246

\bibitem{PT}  Przytycki J., Traczyk P.,
    Proc. Amer. Math. Soc. 100 (1987) 744-748

\bibitem{RT1} Turaev V. G.,
Invent. Math. 92 (1988) 527–533

\bibitem{RT} Reshetikhin N. Yu., Turaev V. G.,
Commun. Math. Phys. 127 (1990) 1–26.

\bibitem{GuadMarMin} Guadagnini E., Martellini M., Mintchev M.,
In: Doebner H.D., Hennig J.D. (eds) Quantum Groups. Lecture Notes in Physics. (1990) 307-317

\bibitem{GuadMarMin2} Guadagnini E., Martellini M., Mintchev M.,
Phys. Lett. B235 (1990) 275

\bibitem{RT2} Mironov A., Morozov A., Morozov An.,
	JHEP03, no. 034. (2012), arXiv:1112.2654

\bibitem{RTend} Itoyama H., Mironov A., Morozov A., Morozov An.,
Int. J. Mod. Phys. A27 (2012) no. 1250099, arXiv:1204.4785

\bibitem{2br} Ramadevi P., Govindarajan T. R., Kaul R. K.,
Mod.Phys.Lett. A9 (1994) 3205-3218, arXiv:hep-th/9401095

\bibitem{2br2} Galakhov D., Melnikov D., Mironov A., Morozov A., Sleptsov A.,
Phys.Lett.B743 (2015) 71-74, arXiv:1412.2616

\bibitem{2br3} Nawata S., Ramadevi P., Singh V. K.,
arXiv:1504.00364

\bibitem{2brend} Mironov A., Morozov A., Morozov An., Ramadevi P., Singh V.K.,
	JHEP. 07 (2015) no. 109, arXiv:1504.00371

\bibitem{Rgates} Kolganov N., Morozov An.,
 	Pis'ma v ZhETF, 111, N9 (2020), arXiv:2004.07764

\bibitem{Rgates2} Kolganov N., Mironov S., Morozov An.,
 	Nuclear Physics B987 (2023) 116072, arXiv:2105.03980

\bibitem{Rgates3} Mironov S., Morozov An., arXiv:2404.12222


\end{thebibliography}

\section{Acknowledgements}

We are grateful to S. Straupe for very useful discussions.
This work was supported by Russian Science Foundation grant No 23-71-10058.


\end{document}